\documentclass[runningheads]{llncs}
\pdfoutput=1
\usepackage[utf8]{inputenc}
\usepackage{listings}
\usepackage{paralist}
\usepackage{multirow}
\usepackage{amsmath}
\usepackage{tabularx}
\usepackage{longtable}
\usepackage{mdframed}

\usepackage{numprint}
\usepackage{listings}
\usepackage{subcaption}
\usepackage{verbatim}

\usepackage[T1]{fontenc}

\title{Ultra-Large Repair Search Space with Automatically Mined Templates: the Cardumen Mode of Astor}
\titlerunning{Ultra-Large Repair Search Space with Automatically Mined Templates}

\usepackage{framed}
\usepackage{pgf}
\usepackage[scale=2]{ccicons}
\usepackage[english]{babel}
\usepackage{hyperref}
\usepackage{algorithm} 
\usepackage{algorithmic}

\newcommand{\ce}[1]{{\bf{#1}}}

\newcommand{\mycode}[1]{\lstinline[basicstyle=\small]$#1$}
\newcolumntype{b}{>{\hsize=1.167\hsize}X}
\newcolumntype{s}{>{\hsize=.5\hsize}X}
\newcommand{\nrpatches}{8935 }
\newcommand{\nronlycardumen}{8 }

\author{Matias Martinez\inst{1}       \and
        Martin Monperrus\inst{2}
}

\institute{  University of Valenciennes, 
              France \\
              \email{matias.martinez@univ-valenciennes.fr}           
              \and
              KTH Royal Institute of Technology, Stockholm, Sweden \\
              \email{martin.monperrus@csc.kth.se}        
}

\begin{document}
\date{Received: date / Accepted: date}
\maketitle

\begin{abstract}
Astor is a program repair library which has different modes.
In this paper, we present the Cardumen mode of Astor, a repair approach based mined templates that has an ultra-large search space. We evaluate the capacity of Cardumen to discover test-suite adequate patches (aka plausible patches) over the 356 real bugs from Defects4J \cite{JustJE2014}.
Cardumen finds \nrpatches patches over 77 bugs of Defects4J. This is the largest number of automatically synthesized patches ever reported, all patches being available in an open-science repository. Moreover, Cardumen identifies \nronlycardumen unique patches, that are patches for Defects4J bugs that were never repaired in the whole history of program repair.
\end{abstract}

\section{Introduction}

There have been major contributions in the field of automatic program repair in recent years.
The program repair community explores different directions, most notably \emph{Generate and Validate (G\&V)} repair approaches \cite{LeGoues2012TSEGP} as well as synthesis-based approaches \cite{Nguyen:2013:SPR,Nopol}.

In this paper, we aim at creating an ultra-large search space, possibly the largest repair search space ever.
To maximize the number of synthesized test-suite adequate patches, we design a new program repair algorithm. 
This algorithm is called Cardumen.
Cardumen extracts code templates  from the code under repair. Those templates contain placeholders to be bound to available variables at a potential repair location. 
Moreover, in order to speed up exploration of the search space, Cardumen uses a probability model for prioritizing candidates patches.

We evaluate the capacity of Cardumen for discovering \emph{test-suite adequate patches} over the 356 real bugs from Defects4J \cite{JustJE2014}.
The results go beyond our initial vision.
First, Cardumen finds \nrpatches patches over 77 bugs of Defects4J. This is the largest number of automatically synthesized patches ever reported for Defects4J. It demonstrates the width of Cardumen's search space.
Second, Cardumen identifies \nronlycardumen unique patches, i.e., patches for Defects4J bugs that were never handled by any system in the whole history of program repair.
This shows that Cardumen's search space is unique, it explores completely uncharted territories of the repair search space.

To sum up, our contributions are:
\begin{enumerate}
\item Cardumen: a novel program repair algorithm that is designed to maximize the number of test-suite adequate patches. It is based on mining repair templates. It uses a novel probabilistic heuristic for prioritizing candidate patches.
\item An analysis of the execution of  Cardumen over 356 real bugs from Defects4J. Cardumen is capable of finding test-suite adequate patches for 77 bugs from Defects4J, including \nronlycardumen uniquely fixed bugs that no other program repair system has ever fixed. For those uniquely fixed bugs, we discuss the unicity of Cardumen's search space.
\item A publicly available list of \nrpatches  test-suite adequate patches for 77 bugs from Defects4J. We envision that this list will support future research in program repair, for instance to improve synthesis of patches, ranking of patches, and dynamic analysis of patches.
\end{enumerate}

The paper is organized as follows:
Section \ref{sec:cardumen} presents the approach Cardumen.
Section \ref{sec:evaluation} evaluates Cardumen over bugs from Defects4J.
Section \ref{sec:relatedwork} presents the related works.
Section \ref{sec:discussion} presents a discussion about the experiment and the threats of validity.
Section \ref{sec:conclusion} concludes the paper.

 \section{Program Repair with Automatically Mined Templates}
\label{sec:cardumen}

We now present the design and the main algorithms of Cardumen.

\subsection{Cardumen in a nutshell}
\label{sec:cardumennutshell}

Cardumen is a repair system designed  for discovering the maximum number of test-adequate patches.
It takes as input: 
\begin{inparaenum}[\it a)]
\item the source code of a buggy program, and 
\item the test-suite of that program with at least one failing test case that exposes the bug.
\end{inparaenum}

Cardumen first applies a spectra fault localization \cite{Reps:1997:UPP} to detect suspicious buggy pieces of code. 
For a given source code location, Cardumen introduces a novel patch synthesis.
The repair always consists of a replacement of the suspicious code element by an instance of a code template.
The code templates are mined and instantiated in a unique two-step process.
The first step is the \emph{Automated Code Template Mining}, which mines code templates from the code of the application under repair for creating a template-based search space (explained in section \ref{sec:templatespace}).
The second step is \emph{Probabilistic-based Template Space Navigation}, 
which uses a probabilistic model for navigating the space of candidate patches, each synthesized from template mined from the application under repair (explained in Section \ref{sec:navigationsearchspace}).

Once a candidate patch is synthesized from a template instantiation, the patched version of the buggy program is executed first on the originally failing test cases and then, if all of them pass, on the remaining test cases (i.e., regression test, which originally pass over the buggy version).

\subsubsection{Example of patch generated by Cardumen}
Cardumen is able to find  test-suite adequate patches for  bug Math-73 from the bug dataset Defects4J \cite{JustJE2014}.
One of them, presented in Listing \ref{listingPatchM73}, modifies the expression of a \ce{return} statement  at line 138 from class \ce{BrentSolver}.

\begin{lstlisting}[caption={Patch for bug Math-73  by Cardumen at BrentSolver class},label={listingPatchM73},basicstyle=\small]
138 - return solve(f, min, yMin, max, yMax, initial, yInitial);
138 + return solve(f, yMin, yMax);
\end{lstlisting}

The template used by Cardumen for synthesizing that patch is presented in Listing \ref{listM73template} and it was mined from the statement \ce{return solve(f, min, max)} written at line 68 of the same class. 

\begin{lstlisting}[caption={Template used for synthesize a patch for bug Math-73}, label={listM73template},basicstyle=\small]
      solve(_UnivariateRealFunction_0, _double_1, _double_2)
\end{lstlisting}

The template has 3 placeholders: the first one \ce{\_UnivariateRealFunction\_0} of type UnivariateRealFunction, the other two  \ce{\_double\_1} and  \ce{\_double\_2} of type Double.
Cardumen creates candidate patches by binding those placeholders with variables available at the repair location.
There, it synthesized 196 patches using the 14 variables of type double and the unique variable of type UnivariateRealFunction available at the line 138.
Cardumen then selected one of them using a probability model which prioritized those patches according to the frequency of the variable names used by each patch.  
For example, the patch from \ref{listingPatchM73}, which uses variables \ce{f, yMin, yMax}, is prioritized before than another patch which uses variables \ce{f, initial, functionValue}, due to the variables of the former patch are used more frequently together in the code than those from the latter patch. 
Finally Cardumen evaluated the selected patch using the test-suite of the buggy program.

\subsection{Cardumen Repair Algorithm}

\begin{algorithm}[t]
\begin{algorithmic}[1]
\REQUIRE{A buggy program $P$}
\REQUIRE{A Test suite $TS$ for $P$ }
\ENSURE{A list of patches $tsa\_patches$ to the buggy program $P$}
    
    \STATE{$suspicious \leftarrow runFaulLocalization(P, TS)$} \label{algo:fl}
     \STATE{$mpl \leftarrow createModifPoint(suspicious)$} \label{algo:mp}
    \STATE{$tsa\_patches \leftarrow \emptyset$} 
    \STATE{$templates \leftarrow mineTemplates(P)$}  \label{algo:minetemp}
    \STATE{$varNameProbModel \leftarrow createVarNameProbModel(P)$}  \label{algo:varprob}
    \STATE{$t \leftarrow 0$}
    \WHILE{$t$ to $MAX\_TIME$} \label{algo:mainloop}
         \STATE{$mp_i \leftarrow chooseMPRandom(mpl)$}  \label{algo:chooseMP}
        \STATE{$t_i \leftarrow chooseTemplateRandom(mp_i)$}  \label{algo:choosetemplate}
        \STATE{$tinstances_i \leftarrow createInstances(mp_i, t_i)$}   \label{algo:createInstances}
        \STATE{$ti_i \leftarrow chooseInstance(tinstances_i, varNameProbModel)$}  \label{algo:chooseInstance}
        \STATE{$pc_i \leftarrow createPatchCode(ti_i)$}  \label{algo:createpatchcode}
        \STATE{$nbFT_i \leftarrow getNbOfFailingTests(TS, P, pc_i$)}  \label{algo:verifypatch}
        \IF{$nbFT_i = 0$} \label{algo:nofailing}
            \STATE{$tsa\_patches \leftarrow tsa\_patches \cup pc_i$}  \label{algo:solution}
        \ENDIF
    \ENDWHILE
\RETURN{$tsa\_patches$}  \label{algo:return}
\end{algorithmic}
\caption{Cardumen's main algorithm}
\label{alg:cardumen}
\end{algorithm}

Algorithm \ref{alg:cardumen} presents the main workflow of Cardumen.
Cardumen first executes a fault localization approach for obtaining the suspicious line (Line \ref{algo:fl}) and it then creates a list of modification points from the suspicious code elements (Line \ref{algo:mp}).
Cardumen proceeds to mine code templates from the application code under repair (Line \ref{algo:minetemp}) and to create the probability model of variables names (Line \ref{algo:varprob}).
After that,  Cardumen starts navigating the search space during an amount of time (Line \ref{algo:mainloop}).
On each step, Cardumen carries out the following steps.
It first randomly selects a modification point $mp_i$ (Line \ref{algo:chooseMP}) and a template (Line \ref{algo:choosetemplate}).
Then, from $mp_i$ and $t_i$, it creates a list of template instances $tinstances_i$ (Line \ref{algo:createInstances}), and from that list, it selects one template instance ($ti_i$) using the probabilistic model (Line \ref{algo:chooseInstance}).
Finally, it synthesizes the patch code from the selected instance (Line \ref{algo:createpatchcode}), and runs the test-suite over the patched application (Line \ref{algo:verifypatch}). If there is not any failing test case (Line \ref{algo:nofailing}), Cardumen adds the patch to a list of test-suite adequate patches (Line \ref{algo:solution}).
At the end, Cardumen returns that list (Line \ref{algo:return}). 

Now, let us describe in detail each step from the mentioned algorithm.

\subsection{Identifying Potential Modification Points}
\label{sec:identifyingmodificationpoints}

The first step carried out by Cardumen is the creation of a representation of the buggy program, which only includes buggy suspicious code elements. This allows to reduce the search space.
Then, Cardumen generates patches by modifying  only those elements, ignoring the rest of the code not included in that representation. 

\subsubsection{Reducing search space using fault localization}
\label{sec:faultlocalization}

For calculating the suspiciouness of code elements,  
Cardumen uses a spectrum based fault localization called GZoltar \cite{gzoltar2012} which produces as output a suspicious value (between 0 and 1) for each statement and method of the program. 
Cardumen first orders decreasing statements according to suspiciousness value and takes the first $X$ statement  with suspicious greater than a given threshold $\gamma$.

\subsubsection{Creation of modification points}
\label{sec:creatingmp}
We call \emph{Modification point} a  code element from the program under repair that could be transformed to synthesize a candidate repair.
Cardumen creates modification points from the filtered suspicious statement returned by the fault localization approach.

Previous approaches on automated work at the level of statements. For example, the original GenProg  \cite{Weimer2009} and its Java implementation contained in the Astor framework  \cite{astor2016} work at that level.  For instance, Astor framework represents each suspicious element with a modification point.
In GenProg, program statements are labeled with the suspicious values and later manipulated by applying operator over them (i.e.,  replacing, removing or adding statements).

In Cardumen, a Modification point is related to fine-grained code elements rather than statements.
Our approach has two main differences w.r.t previous works.
First, it is flexible with respect to the kind of code element that it can manipulate. 
Cardumen receives as input a \emph{set of kinds of code element} to consider as modification points. We call them \emph{Target code types}.
For a code $c$,  its \emph{Target code type} is  the \emph{type} of the AST node root corresponding to  $c$.
For example, the target type of code \ce{a + b} is binary operator.

Second, Cardumen considers the \emph{type} of the object returned by the evaluation of a code element.
For example, the \emph{Return code type} of the expression  \ce{(a or b)}, where \ce{a} and \ce{b} are Boolean variables, is a Boolean, whereas the return type of \ce{(c - d)}, where \ce{c} and \ce{d} are Integer variables, is Integer.

For creating modification points, 
Cardumen receives as input the {Target code types} $tct$  and  the {Return type} $ert$,
then it parses the AST of suspicious code elements,
filtering those AST nodes of types according to sets $tct$ and $ert$, 
and finally creates one modification point $mp$ for each filtered AST node.
By default, the implementation of Cardumen considers expressions as \emph{target code type}, and every kind of object as \emph{return type}.
 
As we will see later, the  \emph{Target code types}  and the \emph{Return type}  are also used for navigating the search space.

Note that, as Cardumen considers fine-grained elements as modification points, it could exist the case that multiples modification points  refer to different code elements included in a single statement.
For example, for the code \ce{(a > b) \&\& ((d - e) > 3)}, Cardumen creates four  modification points: 
one for reference to the whole Boolean expression, the other for the Boolean expression \ce{(a  > b)}, a third one for a Boolean expression \ce{(d - e) > 3}, and the last one for the Integer expression \ce{(d - e)}.

\subsection{Definition of Cardumen's Template-based Search Space}
\label{sec:templatespace}

Once Cardumen has created a list of potential modification points,
it then creates a pool of code templates that are later used to synthesize candidate patches. 

\subsubsection
{Intuition behind the use of code templates}

Cardumen redefines the idea presented by \cite{Weimer2009} and empirically validated by \cite{martinez2014icse} and \cite{Barr2014PSH} which states that the code of a  patch have been already written in the program under repair, they are called the ``repair ingredients'' or ``patch ingredients''.
In GenProg, a candidate patch $c={s_1 , s_2,..,s_n}$ is synthesized from statements taken as-is from somewhere else in the application. 

The core idea of Cardumen is to \emph{reuse code templates} rather than reusing raw, unmodified code elements (such as raw statements in GenProg). 
Contrary to previous works such as PAR \cite{Kim2013} or SPR \cite{spr}, where candidate patches are synthesized from  predefined manually written templates, Cardumen parses the source code of the buggy program and automatically creates the code templates.

\subsubsection
{Mining code templates from source code}
\label{sec:creationtemplates}

For mining templates, Cardumen parses the AST of the program under repair.
For each AST node, Cardumen replaces all variable names by a placeholder composed of the variable type and one numeric identifier.
For example, the code element \ce{((a>b) \&\&  (c>a))}, where  \ce{a}, \ce{b} and \ce{c} are Integer variables, now becomes \ce{((\_int\_1>\_int\_2) \&\& (\_int\_3>\_int\_1))}, where, for instance, \ce{\_int\_1} is a placeholder. 
After the variable renaming, Cardumen obtains a template which is stored in a template pool. Note that Cardumen also stores for a template the \emph{Target code type} and the \emph{return type} (as described in section \ref{sec:creatingmp}): those types take the same value than the \emph{Target code type} and the \emph{Return type} of the code where template was mined.

Cardumen stores each mined template in a structure called \emph{Templates pool}, which is later used when navigating the search space.

\subsection
{Probabilistic-based Navigation of the Code Template Search Space}
\label{sec:navigationsearchspace}

Once Cardumen has created a list of potential modification points and a template pool, it proceeds to navigate the search space for finding test-suite adequate patches.
For synthesizing a patch, Cardumen applies different steps explained in the rest of this section.

\subsubsection{Selecting a modification point}

Cardumen starts the navigation of the search space by selecting one modification point using weighted random selection
The weight of a modification point $mp$ corresponds to the  suspicious value that the fault localization approach assigned to the code pointed by $mp$.

\subsubsection
{Selecting a code template}
\label{sec:templateselection}

Once a modification point $mp_i$ is selected, Cardumen proceeds to select a template that is used for synthesizing candidates patches at $mp_i$.
For that, Cardumen first queries the template pool (defined in Section \ref{sec:creationtemplates}) which returns a list of templates compatible with the suspicious code to be replaced at $mp_i$. Then, Cardumen selects one of of the templates.

Let us dwell on those steps: template pool querying and template selection.
When templates are searched for, the template pool of Cardumen applies two filters: \emph{Compatibility} filter and \emph{Location} filter.

\paragraph{Filtering templates based on compatible types}
\label{sec:filteringcompatiblestemplates}
When selecting a template from the pool, Cardumen must guarantee that the \emph{Return types} of the replacement and of the replaced code \emph{are compatible}.
For example, suppose that a modification point is the expression \ce{(a > b)}. 
The \emph{Return type } of  \ce{(a > b)}  is Boolean.
Cardumen can only replace this expression by an expression  whose return type is also Boolean. Otherwise, the patch will produce an incorrect, uncompilable AST. 

In this example, Cardumen would replace the modification point \ce{(a > b)} by, for example, a template \ce{(isGreater(\_int\_1,\_int\_2))} (with two placeholders \ce{\_int\_1} and \ce{\_int\_2}, whose method invocation \ce{isGreater} returns a Boolean value.

\paragraph{Filtering templates based on code location}
\label{sec:filteringlocatedtemplates}
Cardumen also proposes a mechanism to reduce the number of potential templates  that is based on code location. 
It filters the candidate templates for $mp_i$ according to the location where the template was extracted. 
We call to this filter, the location filter.

Cardumen handles three template location filters, configured by the user: \emph{local}, \emph{package} and \emph{global}.
If the scope filter is set to \emph{local}, Cardumen keeps in the pool all templates mined from code contained in the same file $f$ as the one from where the selected modification point $mp_i$ is located (i.e., $mp_i \, \in \,  f$).
For the \emph{package} scope filter, it keeps all templates deduced from all files of the package containing $mp_i$, 
whereas for the \emph{global} scope filter, Cardumen considers templates deduced from all statements of the program under repair.

\paragraph{Selecting a code template}
\label{sec:selectingtemplate}

For selecting a template from the list of filtered templates, 
Cardumen carries out a weighted random selection, where the probability of selecting a template $t_i$ corresponds to the proportion of code elements that can be represented by $t_i$ (i.e., whose placeholders correspond to the actual values of the expression under consideration).

\subsubsection
{Instantiating a code template}
\label{sec:createInstanceFromTemplate}

Given a template $t_{i}$ and a modification point $mp$, 
a \emph{template instance}  is a binding of each placeholder from $t_{i}$ to a particular variable that are in the scope of $mp$.

The process of \emph{instantiating} a template $t_{i}$ at one location $mp$ consists on finding all template instances, product of the binding of placeholders of the template and variables on the scope of $mp$.
For example, the instantiation of a template with one placeholder $ph$ of type long at a $mp$ with two variables long in scope, $v1$ and $v2$, produces two instances:  one bound $v1$ to $ph$, the other $v2$ to $ph$.
Then, from each instance, Cardumen is able of synthesizing a candidate patch.

\paragraph{Creating template instances for a modification point}
\label{sec:casesinstances}

Given a modification point $mp$ and a template $t_{i}$, 
the template instantiation process has the following steps:
\begin{enumerate}
\item for each placeholder $ph_i$ from the template, Cardumen finds all variables with compatible types from the scope of $mp$, obtaining the set $cv_i={mv_1, mv_2,..., mv_n}$. 
\item if there is no compatible variable for at least one placeholder from the template $t$, i.e.,  $\exists ph_i  | cv_i={\emptyset}$, it means that the template cannot be instantiated at $mp$. 
Thus, the template is discarded and Cardumen continues by selecting another template. We say that $t$ is \emph{sterile} for $mp$. 
\item if all placeholders from the template have compatibles variables, i.e., $\forall ph_i | cv_{i}  \neq \emptyset$, 
Cardumen creates a template instance by choosing, for each placeholder $ph_i$, a compatible variable ${mv_i}$ from  
$cv_i$. 
Hence, a template instance $ti_i$ binds each placeholders to a variable: 
$ti_i = \{(ph_1, mv_{11}), ...,   (ph_n, mv_{n1})\}$, where $n$ is the number of placeholders from template $t$ and  $mv_{ij}$ is a variable that belongs to $cv_i$.
\end{enumerate}

\subsubsection{Prioritizing template instances based on variable names}
\label{sec:probmodel}

The number of template instances for a modification point $mp_i$ and a template $t_j$ is : 
$\prod_{i=1}^{|v|} |cv_{i}| $.
In practice,
this number of instances can be large. 
For example, the instantiation of template \ce{((\_int\_1>\_int\_2) \&\& (\_int\_3>\_int\_4))} at  the place of $mp_i$  with ten integer variables in the scope of $mp_i$, produces 10000 instances (i.e., $10^4$).

With the goal of reducing the search spaces, 
Cardumen prioritizes the template instances based on  variable names as we explain now.

\paragraph{Defining a probabilistic model based on variable name occurrences}
\label{sec:probabilitymodel}

For prioritizing instances, Cardumen automatically creates a binomial distribution model $pml$ to capture the probability $mp$ of $n$ variable names $\{v_1, ..., v_n\}$  to appear together in a statement.

$$pml_n(\{v_1, ..., v_n\}) =  \frac{(number\_statements\_containing\{v_1, ..., v_n\})}{all\_statements\_with\_n\_names}$$,  $\, pml_n  \, \in \, [0,1]$.

In turn, Cardumen defines different models, $pml_i(v_1, ..., v_i)$, 
where each of one captures the probability of occurrence of a set of $i$ variables in a statement with \emph{at least}  $i$ variables. Note that $i \in [1, n]$ where $n$ is the maximum number of variables that a statement (from the program under repair) has. 

For creating the model $pml_n$, Cardumen scans all the statements of the program under repair.
For each statement $s_i$, it collects all variable names:  $vs = \{v_{i1}, ..., v_{in}\}$.
Then,  it updates the model as follows: 
it first creates subsets of variables of size $i$,  ($i  \, \in  \, [1,n]$),
corresponding to all combinations\footnote{Cardumen does not takes in account the order of variable names inside a statement.} of size $i$ that can be created from $vs$.  
Finally, Cardumen updates the model according to each subset.

As example, suppose a model build $pml$ from three statements $s_1$, $s_2$ and $s_3$ composed by the variables 
$v_1 = {a,b,c,x}$, $ v_2 =  {a,b,d}$ and $ v_3 = {a,d,f}$, respectively.
In that model, the probability of having a variable named "a" together with another named "b" is $pml_2(a,b) = 2/3$ ,  and is larger than the probability of having "a" together with "f" ($pml_2(a,f) = 1/3$) . 
As consequence, using that model, for instantiating a template with two placeholders, Cardumen prioritizes an instance with bindings to variables "a" and "b", over another instance with bindings to "a" and "f".

\paragraph{Adding localness to the probability model}
Inspired on the work by Tu el al. \cite{Tu2014LS} about the localness of code, 
which proposes an extended version of n-gram model to capture  local regularities, 
Cardumen creates two sub-models, which conform the probability model $pml$:
one, called `Global' $pml_g$,  which  consider all statements from the program under repair, 
the other, called `Cache' $pml_c$, that only considers  the statements from one file (called Local) or from one package (called Package).
With the same spirit that  \cite{Tu2014LS}, the Global model aims at capturing large global  and static model of variable names, whereas the  cache model aims at modeling a small local (dynamic) name model estimated from the proximate local context (File or Package). 
Consequently, $pml$ is a  linear combination of the two models:

\begin{equation} \label{for:pm}
\begin{aligned}
pm(\{v_1, ..., v_n\})) = \lambda  \cdot pm_g(\{v_1, ..., v_n\}) \\
 + \quad (1 - \lambda) \cdot  pm_c(\{v_1, ..., v_n\}) 
\end{aligned}
\end{equation}

Finally, Cardumen uses the model  $pml$ to obtain the probability of each template instance.
Then, it selects the $\varrho$ instances with higher probability.

\subsubsection{Selecting an instance template}
\label{sec:selectinginstance}

Cardumen selects one instance from the list of instances by applying weighted random selection, where the weight of an instance is given by the probability of its variables' names, calculated using the probability model presented in Section \ref{sec:probmodel}.

\subsubsection
{Synthesizing candidate patch code}
\label{sec:patchsynthesis}

For synthesizing the code of a candidate patch from a template instance, 
Cardumen first  takes the template, creates a clone of it, and replaces each placeholder by the variable bound to it, according to the template instance.
After that, the patch is ready to be applied in the place related to the the modification point. Then, the patched version of the buggy program can be evaluating using the test-suite of the original program.

\subsection{Example: Synthesizing Candidate Patches for Math-70}
\label{sec:exampletemplate}

In this section, we show how Cardumen creates multiple candidate patches for a real-world Java bug included in the bug dataset Defects4J by \cite{JustJE2014}. The subject under study, identified as Math-70, has a bug in class `BisectionSolverImpl'.

Cardumen first identifies 12 modification points (Section \ref{sec:creatingmp}), 10 of them reference statements located on the buggy class `BisectionSolver', the other two reference statements from class `UnivariateRealSolverImpl'. 

Then, Cardumen creates a pool of templates (Section \ref{sec:templatespace}) mined from the application code under repair.
For instance, from the code element: 

\mycode{
if(abs(max - min) <= absoluteAccuracy) 
}
located at line 100 of class \ce{BisectionSolverImpl}, Cardumen mined three templates:
\begin{enumerate}[ 1)]
\item `\mycode{abs((_double_0 - _double_1))) <= (_double_2)}',
of  type ``Binary Operator''  ($<=$) and return type ``Boolean'';
\item  `\mycode{abs( (\_double\_0 - \_double\_1)))}', of type ``Method Invocation'' and Return type ``Double''( mined from left-most term of the $-$ operator);
\item `\mycode{(_double_0 - _double_1)}', of type ``Binary Operator'' ($-$), and Return type ``Double'' (mined from argument of method \mycode{abs}).
\end{enumerate}

For creating a candidate patch, Cardumen first chooses a modification point and a template.
In this example, we suppose that Cardumen first selects: 
\begin{inparaenum}[\it a)]
\item the modification point $mp$ corresponding to the Boolean condition \mycode{(i < maximalIterationCount)} from line 87, which has a suspicious value of $0.5$ (see section \ref{sec:faultlocalization}), and 
\item the template
\mycode{(_double_0 * _double_1) > 0.0}, 
which is a Boolean binary operator mined from line 92 \mycode{if (fm * fmin > 0.0)}  (see section \ref{sec:templateselection}).
\end{inparaenum}

\subsubsection{Instantiating a template}
\begin{table}[t]
\scriptsize
\centering
\caption{Top-5 most frequent var names used in file BisectionSolver (Local) and in the entire buggy application code (Global) from buggy revision  Math-70. }

\begin{tabular}{ |l |l |r| | c | l | }
\hline

\rotatebox[origin=c]{45}{Model}&\rotatebox[origin=c]{45}{\#Vars}&\rotatebox[origin=c]{0}{Variable Names}&\#&\rotatebox[origin=c]{45}{Frequency}\\
\hline
\hline

\multirow{10}{*}{\rotatebox[origin=c]{90}{Local}}
&
\multirow{5}{*}{1}&
min&9&0.069\\
\cline{3-5}
&&upper&8&0.062\\
\cline{3-5}
&&max&8&0.062\\
\cline{3-5}
&&f&8&0.062\\
\cline{2-5}
&
\multirow{5}{*}{2}
&lower, upper&7&0.097\\
\cline{3-5}
&&max, min&7&0.097\\
\cline{3-5}
&& function, lower&3&0.0417\\
\cline{3-5}
&&m, min&3&0.0417\\
\begin{comment}
\cline{2-5}
&
\multirow{5}{*}{3}
&m, max, min&2&0.167\\
\cline{3-5}
&&function, lower, upper&2&0.167\\
\cline{3-5}
&&f, max, min&2&0.167\\
\cline{3-5}
&&initial, lower, upper&2&0.167\\
\cline{3-5}
\cline{2-5}
\end{comment}
&..&..&..&..\\
\hline
\hline
\multirow{10}{*}{\rotatebox[origin=c]{90}{Global}}
&
\multirow{5}{*}{1}
&i&2137&0.058\\
\cline{3-5}
&&length&1591 &0.043\\
\cline{3-5}
&&j&841&0.022\\
\cline{3-5}
&&n&699&0.019\\
\cline{2-5}
&
\multirow{5}{*}{2}
&i, length&365&0.011\\
\cline{3-5}
&&i, j&244&0.007\\
\cline{3-5}
&&data, i&219&0.006\\
\cline{3-5}
&&data, length&199&0.0043\\

\begin{comment}
\cline{2-5}
&
\multirow{5}{*}{3}
&x, y, z&117&0.0051\\
\cline{3-5}
&&data, i, length&86&0.00379\\
\cline{3-5}
&& f, max, min&54&0.00238\\
\cline{2-5}
\end{comment}
&..&..&..&..\\
\hline
  \end{tabular}

\label{tab:frequentvarnames}
\end{table} In the next step, 
Cardumen tries to instantiate 
the selected template by replacing each  of its placeholders (\mycode{_double_0}  and \mycode{_double_1}) by compatible variables that are in the scope at the place of the selected modification point (line 87 of BisectionSolverImpl).
Cardumen found 13 variables of type Double in scope of line 87: 
4 fields on class \textbf{UnivariateRealSolverImpl} (parent class of BisectionSolver),  other 4 fields on \textbf{ConvergingAlgorithmImpl} (parent class of UnivariateRealSolverImpl), 2 parameters for the method \textbf{solves} (which includes line 87), and 3 local variables from that method declared before the line 87.
Using those variables, Cardumen then creates 169 instances of the template obtained from the combination of those variables i.e., $13^2 = 169$. 
For example, a mapping relates the placeholder  \mycode{_double_0} with variable "max" and   \mycode{_double_1} with "min", both variables are parameters of method \textbf{solver}.
After that, Cardumen prioritizes those 169 instances, using a probability model based on variable name frequency (Section \ref{sec:probmodel}).
A portion of this model for subject Math-70 is presented in Table \ref{tab:frequentvarnames}. 
It shows the probabilities of the variable names according to the number of variables per statements (column `\#Vars').
For example, the first row shows that the probability of having a variable named "min" in statements a) with only one variable, and b) from class BisectionSolver (i.e., local model) is 6.9\%.

\subsubsection{Synthesizing the patch}

Cardumen selects one instance using the probability model.
For example,
suppose that Cardumen selects the instance with the mapping between placeholders and variables: (\mycode{_double_0= max}) and (\mycode{_double_1= min}).
Then, Cardumen proceeds to synthesize the candidate patch by replacing the placeholders from the template \mycode{(_double_0 * _double_1) > 0}  by the bound variables given by the instance: \mycode{_double_0}  by "max" and 
placeholder \mycode{double_0} by "min".
This step gives as result the candidate patch \mycode{(max * min) > 0.0}, which can be applied at line 87 of BisectionSolverImpl class.

\subsection{Implementation}
Cardumen is a new mode in the  Astor framework \cite{astor2016} for repairing Java code. Cardumen's implementation uses  Spoon  \cite{spoon} to create the code model of the application under repair. 
For sake of open-science, the source code of Cardumen is publicly available at \url{https://github.com/SpoonLabs/astor}. 
\section{Evaluation}
\label{sec:evaluation}

The research questions that guide the Cardumen evaluation are:

\begin{enumerate}

\item[] RQ 1: To what extent does Cardumen generate test-suite adequate patches?  
\item[]
\item[] RQ 2: 2.a)  Is Cardumen able to 
            identify multiple test-suite adequate patches, i.e., does it have a rich search space?
            2.b) How many bugs can be repaired by a high number of test-suite adequate patches (6 or more patches)? 
            2.c) Does the presence of multiple patches happen often, in several projects?

 \item[]          
\item[] RQ 3: To what extent is Cardumen able to generate a) patches located in different locations, and  b) different kind of patches for a bug?

\end{enumerate}

\begin{table*}
\small

\begin{tabular}{ll}

\begin{tabular}{ |l| l| r| r| r|}
\hline
& Id & \rotatebox[origin=c]{45}{\#Patches} & \#Loc & \rotatebox[origin=c]{45}{\#KindP} \\
\hline
\hline
\multirow{12}{*}{\rotatebox[origin=c]{90}{Closure}}
&Cl7 & 2 &2 & 2 \\
&Cl10 & 4 &2 & 3 \\
&Cl12 & 2 &1 & 1 \\
&Cl13 & 2 &1 & 2 \\
&Cl21 & 135 &3 & 5 \\
&Cl22 & 141 &4 & 5 \\
&Cl33 & 2 &2 & 1 \\
&Cl40 & 4 &1 & 1 \\
&Cl45 & 10 &1 & 3 \\
&Cl46 & 25 &3 & 4 \\
&Cl55 & 1 &1 & 1 \\
&Cl133 & 1 &1 & 1 \\
\cline{2-5}
&12/65 & {\bf 329}  & \multicolumn{2}{c|}{}   \\
\hline
\hline
\multirow{5}{*}{\rotatebox[origin=c]{90}{Time}}
&T4 & 7 &1 & 3 \\
&T7 & 1 &1 & 1 \\
&T9 & 1 &1 & 1 \\
&T11 & 288 &11 & 7 \\
&T17 & 14 &1 & 3 \\
&T18 & 1 &1 & 1 \\
\cline{2-5}
&6/27 & {\bf 312}  &\multicolumn{2}{c|}{}   \\
\hline
\hline
\multirow{7}{*}{\rotatebox[origin=c]{90}{Lang}}
&L7 & 8 &1 & 2 \\
&L10 & 4 &2 & 2 \\
&L14 & 8 &2 & 2 \\
&L22 & 21 &1 & 3 \\
&L24 & 2 &2 & 2 \\
&L27 & 164 &3 & 8 \\
&L39 & 617 &2 & 12 \\
\cline{2-5}
&7/65 & {\bf 824}  & \multicolumn{2}{c|}{}  \\
\hline
\hline
\multirow{12}{*}{\rotatebox[origin=c]{90}{Chart}}
&Ch1 & 780 &1 & 6 \\
&Ch3 & 19 &3 & 3 \\
&Ch4 & 2 &1 & 1 \\
&Ch5 & 700 &6 & 11 \\
&Ch6 & 2 &1 & 1 \\
&Ch7 & 396 &8 & 5 \\
&Ch9 & 23 &4 & 4 \\
&Ch11 & 2 &2 & 1 \\
&Ch12 & 1 &1 & 1 \\
&Ch13 & 1227 &7 & 14 \\
&Ch15 & 24 &1 & 1 \\
&Ch17 & 1 &1 & 1 \\
\hline

\end{tabular}
&
\begin{tabular}{ |l| l| r| r| r|}
\hline
& Id & \rotatebox[origin=c]{45}{\#Patches} & \#Loc & \rotatebox[origin=c]{45}{\#KindP} \\
\hline
\hline
\multirow{4}{*}{\rotatebox[origin=c]{90}{Chart}}
&Ch24 & 3 &1 & 1 \\
&Ch25 & 454 &32 & 25 \\
&Ch26 & 138 &15 & 11 \\
\cline{2-5}
&15/26 & {\bf 3772}  & \multicolumn{2}{c|}{}   \\
\hline
\multirow{28}{*}{\rotatebox[origin=c]{90}{Math}}
&M2 & 28 &5 & 5\\
&M5 & 28 &2 & 3 \\
&M6 & 2 &1 & 1 \\
&M8 & 82 &3 & 4 \\
&M18 & 4 &3 & 3 \\
&M20 & 67 &17 & 12 \\
&M28 & 203 &13 & 14 \\
&M30 & 45 &1 & 2 \\
&M32 & 2 &1 & 2 \\
&M33 & 1 &1 & 1 \\
&M40 & 12 &4 & 5 \\
&M41 & 46 &4 & 2 \\
&M46 & 15 &1 & 1 \\
&M49 & 9 &3 & 3 \\
&M50 & 722 &3 & 10 \\
&M57 & 5 &1 & 1 \\
&M58 & 1 &1 & 1 \\
&M60 & 17 &1 & 2 \\
&M62 & 2 &2 & 2 \\
&M63 & 33 &2 & 4 \\
&M69 & 2 &1 & 1 \\
&M70 & 8 &1 & 1 \\
&M73 & 554 &3 & 3 \\
&M74 & 2 &2 & 2 \\
&M78 & 13 &4 & 5 \\
&M79 & 6 &1 & 1 \\
&M80 & 585 &6 & 12 \\
&M81 & 676 &23 & 16 \\
&M82 & 39 &3 & 3 \\
&M84 & 149 &1 & 4 \\
&M85 & 109 &2 & 5 \\
&M88 & 5 &1 & 1 \\
&M95 & 134 &3 & 11 \\
&M97 & 86 &2 & 3 \\
&M101 & 1 &1 & 1 \\
&M104 & 2 &1 & 1 \\
&M105 & 3 &1 & 1 \\
\cline{2-5}
&37/105 & {\bf 3698}  & \multicolumn{2}{c|}{}   \\
\hline
\multicolumn{2}{|c|}{\bf TOTAL: 77} & {\bf 8935}  & \multicolumn{2}{c|}{}   \\
\hline
\end{tabular}

\end{tabular}
\caption{Identifiers of the  77 bugs from Defects4J repaired by Cardumen, together with the number of different test-suite adequate patches found for each bug (Column \#Patches). Column \#Loc displays the number of different locations the patches are applied. Column  \#KindP displays the number of different kind of expression involved on the patches.}
\label{tab:idrepairedsummary}
\end{table*} 
\subsection{Methodology}
\label{sec:evalmethodology}

We run Cardumen over the Defects4J bug benchmark \cite{JustJE2014}.
Each execution trial is configured as follows. 
Maximum execution time: 3 hours,
maximum number of modification points: 1000 (Section \ref{sec:creatingmp}), 
scope of template ingredients: `package' (Section \ref{sec:filteringlocatedtemplates}), and
maximum number of tried template instances: 1000 (Section \ref{sec:probmodel}).
Since Cardumen is randomized algorithm, we executed 10 trials for each bug from Defects4J.
Note that we do not evaluate Cardumen over bugs from Mockito project included in Defects4J due to a technical issue when parsing the Mockito's code. Bugs from that project were also discarded by the automated repair literature  (e.g., \cite{xin2017leveraging,defects4j-repair,white2017dl,le2016history,Chen2017CPR}).
All the experimental results, including the patches found by Cardumen, are publicly available at  \url{https://github.com/SpoonLabs/astor-experiments/tree/master/cardumen-patches}.

\subsection{RQ 1: To what extent does Cardumen generate test-suite adequate patches?}
\label{sec:repairedbugs}

Table \ref{tab:idrepairedsummary} shows the results of our experiment. It displays the identifier of the bugs from Defects4 repaired by Cardumen (column {Id}), and  the number of unique patches for each bug (column {\#Patches}). The other columns will be explained later.

In total, Cardumen discovers 8935 different test-suite adequate patches for 77 bugs of Defects4J.
Cardumen found one patch (at least) for 15 out of 27  bugs from Chart project,
37 out of 105 for Math, 6 out of 27 for Time, 7 out of 65 for Lang, and 12 out of 135 for Closure.

\begin{framed}
{\bf Response to RQ1:} Cardumen finds {\bf 8935} test-suite adequate patches for {\bf 77} bugs of Defects4J.
\end{framed}

\begin{framed}
Implication for program repair research: So far program repair research has neglected the exploration of the complete search space: most papers report a single patch. However, this experiment shows that the search space is much richer than that. This represents a mine of information on the behavior of the program under repair.
\end{framed}

Additionally, we found that, between those 77 bugs, 
Cardumen is the first repair system to find test-suite adequate patches for {\bf \nronlycardumen} new bugs  of Defects4J, for which no system ever has managed to find a single one.
Those \nronlycardumen uniquely repaired bugs are: 1 bug from Chart (id 11), 3 from Math (ids 62, 101 and 104), 1 from Lang (id 14), 2 from Closure (ids 13 and 46), and 1 from Time (id 9).
For the other 69 bugs repaired by Cardumen, there is at least one other approach that also proposes a test-suite adequate patch.
The repair system that we analyzed where those that:
\begin{inparaenum}[\it 1)]
\item the evaluation was done over the dataset Defects4J; 
\item the identifiers of the repaired bugs from Defect4J are given on the respective paper or included in the appendix.
\end{inparaenum}
They are: ACS \cite{Xiong2017}, Nopol \cite{Nopol,durieux:hal-01480084}, jGenProg \cite{defects4j-repair}, DynaMoth \cite{Durieux2016DDC}, DeepRepair \cite{white2017dl}, GP-FS \cite{gpfl2017}, 
JAID \cite{Chen2017CPR},  ssFix \cite{xin2017leveraging} and HDRepair \cite{le2016history} (for this approach, as  neither the  identifiers of the repaired bugs nor the actual patches were reported, we considered the results reported by ssFix's authors \cite{xin2017leveraging}).

\subsection{Is Cardumen able to identify multiple test-suite adequate patches per bug?}

Now, let us study the number of patches per bug.
Between the 77 patches, 67 of them (87\%) have 2 or more test-suite adequate patches.
We observe that for 32 out of 77 (41.5\%) the number of patches that Cardumen finds is smaller than 5, whereas 10 (13\%) has a single patch.
On the contrary, 19 bugs (24.7\%) can be repaired by more than 100 test-suite adequate patches, and even one bug (Chart-13) has 1227 patches.

\begin{framed}
{\bf Response to RQ 2:}
The results show that:  
{\bf 2.a)} for 67 out of 77 bugs Cardumen found 2+ patches;
{\bf 2.b)} a high abundance of patches occurs frequently (e.g., 45 bugs (58\%) with 6+ patches);  and  
{\bf 2.c)} a high abundance of patches is not project-specific, it is valid to all projects from Defects4J.
\end{framed}

\subsection{RQ 3 (a): To what extent is Cardumen able to generate  patches located in different locations for a bug?}

Each test-suite adequate patch is applied at a specific location (i.e., file name and line).
For each bug, we study the locations of Cardumen's patches.
Column {\#Loc} from Table \ref{tab:idrepairedsummary} displays the number of different locations where the patches are applied.
For instance, bug Chart-11 has two patches, one is applied to class ShapeUtilities at line 274 and the other one is applied to in the same class at line 275.

For 36 out of 77 (46.7\%) bugs, the patch are all applied in a single location.
For 41 (53.3\%) bugs, the Cardumen test-adequate patches are applied to different locations of the buggy application (2+), whereas for 11 out of 77 (14\%) bugs, the number of locations is 5+.
For them, the number of patches is always high (+50).
However, abundance does not depend on number of locations: there are bugs with low number of locations (i.e., 3 or less) but with a large number of patches (Closure-21, Lang-39, Chart-1 and Math-73).

\begin{framed}
{\bf Response to RQ 3: a)}
The results show that Cardumen has the ability to discover 
patches applied at different locations of the buggy application.
This happens for 53\% of the repaired bugs,

\end{framed}

\begin{framed}
Implication for program repair research: The program repair search space is a combination of the location space and the modification space at a given location. This is known, but nobody has ever reported on the actual number of different locations, and we are the first to do so at this scale. Comparing the behavior of patches happening at different locations seems very promising: we envision that the patches would have different execution traces that could be compared one against the other. 
\end{framed}

\subsection{RQ 3 (b): To what extent is Cardumen able to generate different kind of patches for a bug?}

Cardumen has the ability to synthesize patches at the level of expression.
We now study the kinds of expressions involved in each patch to know whether Cardumen is able to synthesize patches that are fundamentally different.

We define the \emph{kind of a patch} as the concatenation of a) the kind of expression of the patch, with b) the kind of the parent element where that expression is applied.  
For example, Math-32 has two test-adequate patches, both replacing the right size of a variable initialization.
The first one, replaces it by a method invocation (\ce{FastMath.max}), the second one by a binary expression (\ce{x * x}).
The kind of expression introduced by the patch are different: the first patch replaces the buggy code by an expression of kind "Method invocation", the second one by another kind of expression: Binary Operator \ce{(*, i.e., multiplication)}.
Then, the parent element of both method invocation (first patch) and binary operator (second patch) is a variable declaration. 
Consequently, the kinds of patches of Math-32 are "Method\_Invocation|LocalVariableDeclaration" and "BinaryOperator|LocalVariableDeclaration".

Column {\#KindP} from Table \ref{tab:idrepairedsummary} gives the number of different kinds of patches per bug.
For 50 out of 77 bugs (65\%), Cardumen found patches with different kinds.
Math-18 is one of those bugs. Cardumen found 4 patches: 2 correspond to a change in a \ce{for} condition, one a change in a \ce{if} condition, and the last a change in right side of an assignment.
For 11 bugs (14\%), the number of different kinds involved in the patch is 10 or more.

The remaining 27 out of 77 bugs (35\%) have patches that all involve the same kind of patch.
For instance, Math-6 has 2 patches, both applied to the same location, which replace a buggy method invocation inside a return statement, but those invocations are different (be the message or the called object).

\begin{framed}
{\bf Response to RQ 3: b)}
For the majority of the repaired bugs (65\%), Cardumen found test-suite adequate patches whose 
kinds are different, the patches are made over different kinds of code elements.
This shows the richness and variety of Cardumen's repair search space.
\end{framed}

\begin{framed}
Implication for program repair research: So far, program repair has mostly focused a handful of specific kind of patches (e.g., conditions or RHS of assignments). The open-ended search space of Cardumen enables the community to identify novel kinds of patches for which dedicated repair algorithms will eventually be devised in the future.  
\end{framed}

\section{Related Work}
\label{sec:relatedwork}
\subsection{Repair Approaches}

\subsubsection{Test-suite based repair approaches}
One of the most popular families of automated program repair recently proposed are 
\emph{Generate-and-validate repair techniques}. 
Those kind of techniques first search within a search space to generate a set of patches, and then validate the generated patches. The \emph{Test-suite based repair approach} family uses test-suites for validating the generated patches.
GenProg  \cite{Weimer2009,LeGoues2012TSEGP}, one of the earliest generate-and-validate techniques, uses genetic programming to search the repair space and generates patches created from existing code from elsewhere in the same program. It has three repair operators: add, replace or remove statements.
Other approaches have extended GenProg: for example, AE \cite{weimer2013AE} employs a novel deterministic search strategy and uses program equivalence relation to reduce the patch search space.
RSRepair \cite{rsrepair} has the same search space as GenProg but uses random search instead, and the empirical evaluation shows that random search can be as effective as genetic programming.
The original implementation of GenProg \cite{Weimer2009} targets C code and was evaluated against dataset with C bugs such as ManyBugs and IntroClass \cite{LeGoues2015MB}.
It exists other implementations of GenProg for targeting other code languages, for example, jGenProg, built over the framework Astor \cite{astor2016}, is an implementation of the approach in Java language that targets Java bugs. 
Wen et al. \cite{gpfl2017} presented a systematic empirical study that explores the influence of fault space on search-based repair techniques. 
For the experiment, they created GP-FS, a modified GenProg (i.e., the java implementation jGenProg \cite{astor2016}) which receives as input a faulty space. In their experiment, the authors generated several fault spaces with different accuracy, and then they feed GenProg with those spaces,  finding that GP-FS is capable of fixing more bugs correctly when fault spaces with high accuracy are fed. 

Cardumen has two main differences with respect to those approaches.
The first one is it works at a fine-grained level rather than statements: Cardumen is able to repair expressions insides a statement.
The second is the use of templates derived from the program under repair, rather than the reuse of statements without applying any modification.

The approach ACS (Automated Condition Synthesis) \cite{Xiong2017}, targets to insert or modify an “if” condition to repair defects. ACS  combines three heuristic ranking techniques that exploit 1) the structure of the buggy program, 2) the document of the buggy program (i.e., Javadoc comments embedded in the source code), and 3) the conditional expressions in existing projects.
NpeFix \cite{SANER2017} focuses on repairing null-pointer exceptions.

Contrary to them, Cardumen targets to any kind of code elements (due to its works at the expression level) rather than to a particular defect case (such as ``If'' conditions for ACS).

\subsubsection{Template based repair approaches}

Other approaches have proposed new set of  repair operators.
PAR \cite{Kim2013}, which shares the same search strategy with GenProg, uses  patch templates derived from human-written patches to construct the search space.
SPR  \cite{spr} uses a set of predefined transformation schemas to construct the search space, and patches are generated by instantiating the schemas with condition synthesis techniques. 
JAID \cite{Chen2017CPR} is a state-based dynamic program analyses
which synthesizes patches based on schemas (5 in total).
Each schema triggers a fix action when a suspicious state in the system is reached during a computation. JAID has 4 types of fix actions, such as modify the state directly by assignment, and affects the state that is used in an expression.
Contrary to them,  Cardumen does not have neither any predefined transformation schema nor template: it automatically mines them from the application under repair.

\subsubsection{Approaches guided by examples}

There are approaches that leverage on human written bug fixes. For example, 
Genesis \cite{Long2017AIC} automatically infers code transforms for automatic patch generation. The code transformation used Genesis are automatically infer from previous successful patches.
HRD \cite{le2016history} leverages on the development history to effectively guide and drive a program repair process.
The approach first mines bug fix patterns from the history of many projects and  then employs existing mutation operators to generate fix candidates for a given buggy program.
Both approaches need as input, in addition to the buggy program and its test suite, a set of bug fixes.
Two approaches leveraged on semantics-based examples.
SearchRepair \cite{Ke2015RPS} uses a large database of human-written code fragments encore as satisfiability modulo theories (SMT) constraints on their input-output behavior for synthesizing candidates repairs.
S3 (Syntax- and Semantic-Guided Repair Synthesis)  \cite{Le2017SSS}, a repair synthesis engine that leverages programming-by-examples methodology to synthesize repairs. 
Contrary to them,  Cardumen does not use any extra information rather than the buggy program code and its test-suite: it deduces the templates \emph{on-the-fly}, (i.e., during the repair of a give buggy program)  from the code of the application under repair.

The approach ssFix \cite{xin2017leveraging} performs syntactic code search to find existing code from a code database (composed by the application under repair and external applications) that is syntax-related to the context of a bug statement. 
The approach applies code transformation to adapt the selected code existing code into the buggy location, leveraging the candidate patch. 
Contrary or it, Cardumen leverages on templates mined from the application under repair and does not transform the template code: it binds template placeholders with variables from the context of the buggy statement.

\subsubsection{Probabilistic models based repair approaches}
As Cardumen, there are other approaches that leverage on probabilistic model.
An extension of SPR, Prophet \cite{prophet} applies probabilistic models of correct code learned from successful human patches to prioritize candidate patches so that the correct patches could have higher rankings. 
DeepRepair \cite{white2017dl}, an extension of jGenProg, which navigates the patch search space guided by method and class similarity measures  inferred deep
unsupervised learning.
Martinez et Monperrus \cite{Martinez2013} proposed to probabilistic model built from bug fixes to guide the navigation of the search space.
Contrary to those works, Cardumen builds the probability model from the code under repair, without leverage on provided human bug fixes.

\subsection{Patches analysis}

Recent studies have analyzed the patches generated by some of the approaches we listed before.
The results of those studies show that generated patches may just overfit the available test cases, meaning that they will break untested but desired functionality. 
For example, Qi et al. \cite{Qi2015} find that the vast majority of patches produced by GenProg, RSRepair, and AE avoid bugs simply by functionality deletion. 
A subsequent study by Smith et al. \cite{smith2015cure} further confirms that the patches generated by of GenProg and RSRepair fail to generalize. 
An empirical study \cite{defects4j-repair} reveals that among the 47 bugs fixed by jGenProg, jKali, and Nopol, only 9 bugs are correctly fixed, the rest being overfitting. 
Jiang et al. \cite{jiang2017can} analyzed the Defects4J dataset for finding bugs with weak test cases.
They results shows that 42 (84\%) of the 50 defects could be fixed
with weak test suites, indicating that, beyond the current
techniques have a lot of rooms for improvement, weak test suites may
not be the key limiting factor for current techniques.

\subsection{Analysis of repair search spaces}

Long et al. \cite{Long2016ASS} presented  a systematic analysis of the SPR and Prophet search spaces. The analysis focused on the density of correct and plausible patches in the search spaces, on the ability of those approaches to prioritize correct patches. Some of the finding were: 
the relatively abundant plausible (i.e., overfitted test-adequate) patches in the search space compare to the correct, sparse correct patches, and the effectiveness of both SPR and Prophet at isolating correct patches within the explored plausible patches.

Weimer et al. \cite{weimer2013AE} presented an study of the size of the search space  considered by AE and GenProg approaches. Their goal was to compare the improvement introduced by AE (such as program equivalence) over GenProg.
Their results shows and that  AE dramatically reduces the search space by 88\%, when compared with GenProg and, at the same time, keeps the same repair effectiveness than GenProg.

\subsection{Repair approaches extension for avoiding overfitted patches}

Due to the problematic of test overfitting, recent works \cite{Liu2017IPC,Zu2017Test4Repair} propose to extend existing automated repair approach such as Nopol, ACS and jGenProg.
Those extended approaches generate new test inputs to enhance the test suites and use their behavior similarity to determine patch correctness.
For example, Lui et al \cite{Liu2017IPC} reported that their approach, based on patch and test similarity analysis,  successfully prevented 56.3\% of the incorrect patches to be generated, without blocking any correct patches.
Yang et al. \cite{Yang2017BTC} presented a framework named Opad (Overfitted PAtch Detection) to detect overfilled patches by enhancing existing test cases  using fuzz testing and employing two new test oracles. Opad  filters out 75.2\% (321/427) overfitted patches generated by GenProg/AE, Kali, and SPR.

 \section{Discussion}
\label{sec:discussion}

\subsection{Threats to validity}

\paragraph{Internal threats:}
Due to Cardumen being stochastic, we executed Cardumen over each bugs 10 times for 3 hour, each trial with a different seed. In total, our evaluation took approximately 10710 hours of execution 
equivalent to 446 days.\footnote{Total execution time: $10710 \quad = \quad 357 \quad  bugs \quad  X\quad  10 \quad  trials \quad  X\quad  3\quad  hs$}
Running more executions will involve that Cardumen navigates places from the search space not yet visited and thus potentially  discovers new patches.
Moreover, the experimental setup could impact on the repairability, for instance, we decide to consider the 1000 most suspicious modification points. Between the excluded modification points it could exist one or more places where Cardumen could generate a test-suite adequate patch.

\paragraph{External threats:}
We run Cardumen over 356 bugs from 5 open-source Java projects. 
More bugs from other kind of applications (all evaluated are libraries) could help to validate the efficacy of Cardumen.
As studied by \cite{Qi2015,defects4j-repair}, test-suite based repair approaches can generate plausible patches, yet incorrect. That means, they pass all the test cases from a suite but they are incorrect due to the limitation of the bug oracle: when using test suite as oracle, this limitation is the missing of inputs and outputs for correctly exercising the patch.
Currently,  approaches by \cite{Xin:2017:ITP,Zu2017Test4Repair} aim at improving the quality of test suite for avoiding accepting incorrect patches.
However, in this work, we do not focus on the correctness of patches, which demands another correctness oracle, yet manual or automated: our goal is Cardumen finds the most quantity of code changes that produce a buggy version of a program passing a test-suite (either the original test suite or one augmented). 

\subsection{Limitations}

As described previously in Section \ref{sec:cardumennutshell}, Cardumen synthesizes patches that only modify expressions.
Thus, Cardumen is not able to synthesize patches that add new statement or remove code.
However, we believe that each repair approach focuses on particular defect classes.
We envision that the general process of repair a bug automatically is composed by the execution of different approaches, each targeting on particular set of defect classes.
We aim at implement that vision in our repair framework Astor, which already includes different repair approaches such as jGenProg, jKali, jMutRepair (an implementation of approached proposed by \cite{debroy2010using}), DeepRepair and, from now, Cardumen.
 
\section{Conclusion}
\label{sec:conclusion}

In this paper, we take an original approach to program repair: we aim at finding as many test-suite adequate patches as possible for a given bug.
For that, we created an automated repair approach named Cardumen.
Cardumen found in total \nrpatches test-suite adequate patches, repairing 77 bugs from Defects4J, \nronlycardumen of them not previously repaired by any other repair system.
This result shows the richness of Cardumen's search space. Furthermore, 53\% of repaired bugs have patches applied on different locations; and  65\% of the repaired bugs have different kinds of patches.

As future work, we envision a new repair system that would perform:
\begin{inparaenum}[\it 1)]
\item a first reduction of the complete search space to a subspace only composed of \emph{test-adequate patches}; 
and 
\item a second reduction of that space to a subspace with only \emph{correct} patches.
For implementing this approach, our future plan is to study and compare the \nrpatches patches generated by Cardumen.
\end{inparaenum} 
\bibliographystyle{plain}
\bibliography{references}

\begin{thebibliography}{10}

\bibitem{Barr2014PSH}
Earl~T. Barr, Yuriy Brun, Premkumar Devanbu, Mark Harman, and Federica Sarro.
\newblock The plastic surgery hypothesis.
\newblock In {\em Proceedings of the 22Nd ACM SIGSOFT International Symposium
  on Foundations of Software Engineering}, FSE 2014, pages 306--317, New York,
  NY, USA, 2014. ACM.

\bibitem{gzoltar2012}
J.~Campos, A.~Riboira, A.~Perez, and R.~Abreu.
\newblock Gzoltar: an eclipse plug-in for testing and debugging.
\newblock In {\em 2012 Proceedings of the 27th IEEE/ACM International
  Conference on Automated Software Engineering}, pages 378--381, Sept 2012.

\bibitem{Chen2017CPR}
Liushan Chen, Yu~Pei, and Carlo~A. Furia.
\newblock Contract-based program repair without the contracts.
\newblock In {\em Proceedings of the 32Nd IEEE/ACM International Conference on
  Automated Software Engineering}, ASE 2017, pages 637--647, Piscataway, NJ,
  USA, 2017. IEEE Press.

\bibitem{debroy2010using}
Vidroha Debroy and W.~Eric Wong.
\newblock Using mutation to automatically suggest fixes for faulty programs.
\newblock In {\em Proceedings of the 2010 Third International Conference on
  Software Testing, Verification and Validation}, ICST '10, pages 65--74, 2010.

\bibitem{SANER2017}
T.~Durieux, B.~Cornu, L.~Seinturier, and M.~Monperrus.
\newblock Dynamic patch generation for null pointer exceptions using
  metaprogramming.
\newblock In {\em 2017 IEEE 24th International Conference on Software Analysis,
  Evolution and Reengineering (SANER)}, pages 349--358, Feb 2017.

\bibitem{durieux:hal-01480084}
Thomas Durieux, Benjamin Danglot, Zhongxing Yu, Matias Martinez, Simon Urli,
  and Martin Monperrus.
\newblock {The Patches of the Nopol Automatic Repair System on the Bugs of
  Defects4J version 1.1.0}.
\newblock Research Report hal-01480084, {Universit{\'e} Lille 1 - Sciences et
  Technologies}, 2017.

\bibitem{Durieux2016DDC}
Thomas Durieux and Martin Monperrus.
\newblock Dynamoth: Dynamic code synthesis for automatic program repair.
\newblock In {\em Proceedings of the 11th International Workshop on Automation
  of Software Test}, AST '16, pages 85--91, New York, NY, USA, 2016. ACM.

\bibitem{LeGoues2015MB}
C.~Le Goues, N.~Holtschulte, E.~K. Smith, Y.~Brun, P.~Devanbu, S.~Forrest, and
  W.~Weimer.
\newblock The manybugs and introclass benchmarks for automated repair of c
  programs.
\newblock {\em IEEE Transactions on Software Engineering}, 41(12):1236--1256,
  Dec 2015.

\bibitem{LeGoues2012TSEGP}
C.~Le Goues, T.~Nguyen, S.~Forrest, and W.~Weimer.
\newblock Genprog: A generic method for automatic software repair.
\newblock {\em IEEE Transactions on Software Engineering}, 38(1):54--72, Jan
  2012.

\bibitem{jiang2017can}
Jiajun Jiang and Yingfei Xiong.
\newblock Can defects be fixed with weak test suites? an analysis of 50 defects
  from defects4j.
\newblock {\em arXiv preprint arXiv:1705.04149}, 2017.

\bibitem{JustJE2014}
Ren{\'e} Just, Darioush Jalali, and Michael~D. Ernst.
\newblock {Defects4J}: A database of existing faults to enable controlled
  testing studies for {J}ava programs.
\newblock In {\em Proceedings of the International Symposium on Software
  Testing and Analysis (ISSTA)}, pages 437--440, San Jose, CA, USA, July~23--25
  2014.

\bibitem{Ke2015RPS}
Yalin Ke, Kathryn~T. Stolee, Claire~Le Goues, and Yuriy Brun.
\newblock Repairing programs with semantic code search (t).
\newblock In {\em Proceedings of the 2015 30th IEEE/ACM International
  Conference on Automated Software Engineering (ASE)}, ASE '15, pages 295--306,
  Washington, DC, USA, 2015. IEEE Computer Society.

\bibitem{Kim2013}
Dongsun Kim, Jaechang Nam, Jaewoo Song, and Sunghun Kim.
\newblock Automatic patch generation learned from human-written patches.
\newblock In {\em Proceedings of the 2013 International Conference on Software
  Engineering}, ICSE '13, pages 802--811, Piscataway, NJ, USA, 2013. IEEE
  Press.

\bibitem{Le2017SSS}
Xuan-Bach~D. Le, Duc-Hiep Chu, David Lo, Claire Le~Goues, and Willem Visser.
\newblock S3: Syntax- and semantic-guided repair synthesis via programming by
  examples.
\newblock In {\em Proceedings of the 2017 11th Joint Meeting on Foundations of
  Software Engineering}, ESEC/FSE 2017, pages 593--604, New York, NY, USA,
  2017. ACM.

\bibitem{le2016history}
Xuan Bach~D Le, David Lo, and Claire Le~Goues.
\newblock History driven program repair.
\newblock In {\em Software Analysis, Evolution, and Reengineering (SANER), 2016
  IEEE 23rd International Conference on}, volume~1, pages 213--224. IEEE, 2016.

\bibitem{Liu2017IPC}
Xinyuan Liu, Muhan Zeng, Yingfei Xiong, Lu~Zhang, and Gang Huang.
\newblock Identifying patch correctness in test-based automatic program repair,
  2017.

\bibitem{Long2017AIC}
Fan Long, Peter Amidon, and Martin Rinard.
\newblock Automatic inference of code transforms for patch generation.
\newblock In {\em Proceedings of the 2017 11th Joint Meeting on Foundations of
  Software Engineering}, ESEC/FSE 2017, pages 727--739, New York, NY, USA,
  2017. ACM.

\bibitem{spr}
Fan Long and Martin Rinard.
\newblock Staged program repair with condition synthesis.
\newblock In {\em Proceedings of the 2015 10th Joint Meeting on Foundations of
  Software Engineering}, ESEC/FSE 2015, pages 166--178, New York, NY, USA,
  2015. ACM.

\bibitem{Long2016ASS}
Fan Long and Martin Rinard.
\newblock An analysis of the search spaces for generate and validate patch
  generation systems.
\newblock In {\em Proceedings of the 38th International Conference on Software
  Engineering}, ICSE '16, pages 702--713, New York, NY, USA, 2016. ACM.

\bibitem{prophet}
Fan Long and Martin Rinard.
\newblock Automatic patch generation by learning correct code.
\newblock {\em SIGPLAN Not.}, 51(1):298--312, January 2016.

\bibitem{defects4j-repair}
Matias Martinez, Thomas Durieux, Romain Sommerard, Jifeng Xuan, and Martin
  Monperrus.
\newblock Automatic repair of real bugs in java: A large-scale experiment on
  the defects4j dataset.
\newblock {\em Empirical Software Engineering}, pages 1--29, 2016.

\bibitem{Martinez2013}
Matias Martinez and Martin Monperrus.
\newblock Mining software repair models for reasoning on the search space of
  automated program fixing.
\newblock {\em Empirical Software Engineering}, pages 1--30, 2013.

\bibitem{astor2016}
Matias Martinez and Martin Monperrus.
\newblock Astor: A program repair library for java (demo).
\newblock In {\em Proceedings of the 25th International Symposium on Software
  Testing and Analysis}, ISSTA 2016, pages 441--444, New York, NY, USA, 2016.
  ACM.

\bibitem{martinez2014icse}
Matias Martinez, Westley Weimer, and Martin Monperrus.
\newblock Do the fix ingredients already exist? an empirical inquiry into the
  redundancy assumptions of program repair approaches.
\newblock In {\em Companion Proceedings of the 36th International Conference on
  Software Engineering}, ICSE Companion 2014, pages 492--495, 2014.

\bibitem{Nguyen:2013:SPR}
Hoang Duong~Thien Nguyen, Dawei Qi, Abhik Roychoudhury, and Satish Chandra.
\newblock Semfix: Program repair via semantic analysis.
\newblock In {\em Proceedings of the 2013 International Conference on Software
  Engineering}, ICSE '13, pages 772--781, Piscataway, NJ, USA, 2013. IEEE
  Press.

\bibitem{spoon}
Renaud Pawlak, Martin Monperrus, Nicolas Petitprez, Carlos Noguera, and Lionel
  Seinturier.
\newblock Spoon: A library for implementing analyses and transformations of
  java source code.
\newblock {\em Software: Practice and Experience}, page~na, 2015.

\bibitem{rsrepair}
Yuhua Qi, Xiaoguang Mao, Yan Lei, Ziying Dai, and Chengsong Wang.
\newblock Does genetic programming work well on automated program repair?
\newblock In {\em Computational and Information Sciences (ICCIS), 2013 Fifth
  International Conference on}, pages 1875--1878. IEEE, 2013.

\bibitem{Qi2015}
Zichao Qi, Fan Long, Sara Achour, and Martin Rinard.
\newblock An analysis of patch plausibility and correctness for
  generate-and-validate patch generation systems.
\newblock In {\em Proceedings of the 2015 International Symposium on Software
  Testing and Analysis}, ISSTA 2015, pages 24--36, New York, NY, USA, 2015.
  ACM.

\bibitem{Reps:1997:UPP}
Thomas Reps, Thomas Ball, Manuvir Das, and James Larus.
\newblock The use of program profiling for software maintenance with
  applications to the year 2000 problem.
\newblock In {\em Proceedings of the 6th European SOFTWARE ENGINEERING
  Conference Held Jointly with the 5th ACM SIGSOFT International Symposium on
  Foundations of Software Engineering}, ESEC '97/FSE-5, pages 432--449, New
  York, NY, USA, 1997. Springer-Verlag New York, Inc.

\bibitem{smith2015cure}
Edward~K Smith, Earl~T Barr, Claire Le~Goues, and Yuriy Brun.
\newblock Is the cure worse than the disease? overfitting in automated program
  repair.
\newblock In {\em Proceedings of the 2015 10th Joint Meeting on Foundations of
  Software Engineering}, pages 532--543. ACM, 2015.

\bibitem{Tu2014LS}
Zhaopeng Tu, Zhendong Su, and Premkumar Devanbu.
\newblock On the localness of software.
\newblock In {\em Proceedings of the 22Nd ACM SIGSOFT International Symposium
  on Foundations of Software Engineering}, FSE 2014, pages 269--280, New York,
  NY, USA, 2014. ACM.

\bibitem{weimer2013AE}
W.~Weimer, Z.P. Fry, and S.~Forrest.
\newblock Leveraging program equivalence for adaptive program repair: Models
  and first results.
\newblock In {\em Automated Software Engineering (ASE), 2013 IEEE/ACM 28th
  International Conference on}, pages 356--366, Nov 2013.

\bibitem{Weimer2009}
Westley Weimer, ThanhVu Nguyen, Claire Le~Goues, and Stephanie Forrest.
\newblock Automatically finding patches using genetic programming.
\newblock In {\em Proceedings of the 31st International Conference on Software
  Engineering}, ICSE '09, pages 364--374, 2009.

\bibitem{gpfl2017}
Ming Wen, Junjie Chen, Rongxin Wu, Dan Hao, and Shing-Chi Cheung.
\newblock An empirical analysis of the influence of fault space on search-based
  automated program repair, 2017.

\bibitem{white2017dl}
Martin White, Michele Tufano, Matias Martinez, Martin Monperrus, and Denys
  Poshyvanyk.
\newblock Sorting and transforming program repair ingredients via deep learning
  code similarities, 2017.

\bibitem{Xin:2017:ITP}
Qi~Xin and Steven~P. Reiss.
\newblock Identifying test-suite-overfitted patches through test case
  generation.
\newblock In {\em Proceedings of the 26th ACM SIGSOFT International Symposium
  on Software Testing and Analysis}, ISSTA 2017, pages 226--236, New York, NY,
  USA, 2017. ACM.

\bibitem{xin2017leveraging}
Qi~Xin and Steven~P Reiss.
\newblock Leveraging syntax-related code for automated program repair.
\newblock In {\em Proceedings of the 32nd IEEE/ACM International Conference on
  Automated Software Engineering (ASE)}, pages 660--670. IEEE, 2017.

\bibitem{Xiong2017}
Yingfei Xiong, Jie Wang, Runfa Yan, Jiachen Zhang, Shi Han, Gang Huang, and
  Lu~Zhang.
\newblock Precise condition synthesis for program repair.
\newblock In {\em Proceedings of the 39th International Conference on Software
  Engineering}, ICSE '17, pages 416--426, Piscataway, NJ, USA, 2017. IEEE
  Press.

\bibitem{Nopol}
Jifeng Xuan, Matias Martinez, Favio Demarco, Maxime Clément, Sebastian
  Lamelas, Thomas Durieux, Daniel Le~Berre, and Martin Monperrus.
\newblock Nopol: Automatic repair of conditional statement bugs in java
  programs.
\newblock {\em IEEE Transactions on Software Engineering}, 2016.

\bibitem{Yang2017BTC}
Jinqiu Yang, Alexey Zhikhartsev, Yuefei Liu, and Lin Tan.
\newblock Better test cases for better automated program repair.
\newblock In {\em Proceedings of the 2017 11th Joint Meeting on Foundations of
  Software Engineering}, ESEC/FSE 2017, pages 831--841, New York, NY, USA,
  2017. ACM.

\bibitem{Zu2017Test4Repair}
Zhongxing Yu, Matias Martinez, Benjamin Danglot, Thomas Durieux, and Martin
  Monperrus.
\newblock Alleviating patch overfitting with automatic test generation: a study
  of feasibility and effectiveness for the nopol repair system.
\newblock {\em Empirical Software Engineering}, May 2018.

\end{thebibliography}

\end{document}